\documentclass[sigconf, nonacm]{acmart}

\usepackage{pgfplots}
\usepackage{pgfplotstable}
\pgfplotsset{compat=1.5,scale only axis}

\usepackage{graphicx}
\usepackage{balance}  
\usepackage{multirow}
\usepackage{url}
\usepackage{xcolor}
\usepackage{listings}
\usepackage{caption}
\usepackage{subcaption}
\usepackage{enumitem}
\usepackage{hyperref} 
\usepackage{comment}

\usepackage[para]{footmisc}

\usepackage{booktabs} 

\usepackage[ruled]{algorithm2e}

\newcommand{\para}[1]{\vspace{1mm}\noindent\textbf{#1.}}

\newcounter{AsteriosNOC}
\stepcounter{AsteriosNOC}

\newcounter{MariosNOC}
\stepcounter{MariosNOC}

\AtBeginDocument{%
  \providecommand\BibTeX{{%
    \normalfont B\kern-0.5em{\scshape i\kern-0.25em b}\kern-0.8em\TeX}}}

\definecolor{n1_code}{HTML}{405063}
\definecolor{props}{HTML}{db597c}

\definecolor{gray}{rgb}{0.5,0.5,0.5}
\definecolor{lg}{HTML}{DDDDDD}
\definecolor{CC}{HTML}{666666}
\definecolor{numbers}{HTML}{444444}

\definecolor{bgreen}{HTML}{20BC5F}

\lstdefinelanguage{java}{
  morekeywords={abstract,case,catch,class,def,%
    do,else,extends,false,final,finally,toMatrix,%
    for,if,implicit,import,match,mixin,%
    new,null,object,override,package,%
    private,protected,requires,return,sealed,%
    super,this,throw,trait,true,try,%
    type,val,var,while,with,yield,
    flatMap, filter,aggregate,writeTo,readFrom,groupingKey,Pipeline,
    BatchStage,StreamStage, hashJoin
    }, 
  otherkeywords={=>,<-,<\%,<:,>:,\#,@},
  sensitive=true,
  morecomment=[l]{//},
  morecomment=[n]{/*}{*/},
  morestring=[b]",
  morestring=[b]',
  morestring=[b]"""
}

\lstdefinestyle{javalang}{
  frame=none,
  language=java,
  showstringspaces=false,
  columns=flexible,
  basicstyle={\footnotesize\ttfamily},
  numbers=none,
  numberstyle=\tiny\color{numbers},
  numbersep=5pt,
  keywordstyle=\bfseries\color{n1_code},
  commentstyle=\color{CC},
  stringstyle=\bfseries,
  breaklines=false,
  breakatwhitespace=false,
  tabsize=2,
  xleftmargin=.0in,
  captionpos=b,
  keepspaces=true,
  escapechar=|
}

\lstdefinestyle{javainline}{
  frame=none,
  language=java,
  showstringspaces=false,
  columns=flexible,
  basicstyle={\ttfamily},
  numbers=none,
  numberstyle=\tiny\color{gray},
  keywordstyle=\bfseries\color{n1_code},
  commentstyle=\color{gray},
  stringstyle=\bfseries,
  breaklines=false,
  breakatwhitespace=false,
  tabsize=2,
  xleftmargin=.1in,
  captionpos=b,
  keepspaces=true,
  escapechar=|,
  moredelim=**[is][\color{n1_code}]{`}{`},
  moredelim=**[is][\color{props}]{~}{~},
}

\newcommand\blfootnote[1]{%
  \begingroup
  \renewcommand\thefootnote{}\footnote{#1}%
  \addtocounter{footnote}{-1}%
  \endgroup
}

\begin{document}

\title{Hazelcast Jet: Low-latency Stream Processing at the 99.99$^{\text{th}}$ Percentile}

\author{Can Gencer \hfill Marko Topolnik \hfill Viliam Ďurina \hfill Emin Demirci$^{\ddagger,*}$ \hfill Ensar B. Kahveci$^{\dagger,*}$ 
\hfill Ali Gürbüz Ondřej Lukáš \hfill József Bartók \hfill Grzegorz Gierlach \hfill František Hartman \hfill Ufuk Yılmaz \\ Mehmet Doğan$^{\ddagger,*}$ \hfill Mohamed Mandouh$^{\epsilon,*}$ \hfill Marios Fragkoulis$^{\delta,*}$ \hfill Asterios Katsifodimos$^{\delta,*}$}

\affiliation{%
\vspace{2mm}
  \institution{{\color{white}-} \hfill Hazelcast Inc. \hfill $^\delta$TU Delft \hfill   $^\epsilon$Mansoura University   \hfill $^\ddagger$Layer Co. \hfill {\color{white}-}}
}

\renewcommand{\shortauthors}{Gencer, Topolnik, et al.}

\begin{abstract}
Jet is an open-source, high-performance, distributed stream processor built at Hazelcast during the last five years. Jet was engineered with millisecond latency on the 99.99th percentile as its primary design goal. Originally Jet's purpose was to be an execution engine that performs complex business logic on top of streams generated by Hazelcast's In-memory Data Grid (IMDG):  a set of high-performance, in-memory, partitioned and replicated data structures. With time, Jet evolved into a full-fledged, scale-out stream processor that can handle out-of-order streams and exactly-once processing guarantees. Jet's end-to-end latency lies in the order of milliseconds, and its throughput in the order of millions of events per CPU-core. This paper presents main design decisions we made in order to maximize the performance per CPU-core, alongside lessons learned, and an empirical performance evaluation.
\end{abstract}




\maketitle

\begin{figure}[t]
\begin{center}
\includegraphics[width=\columnwidth]{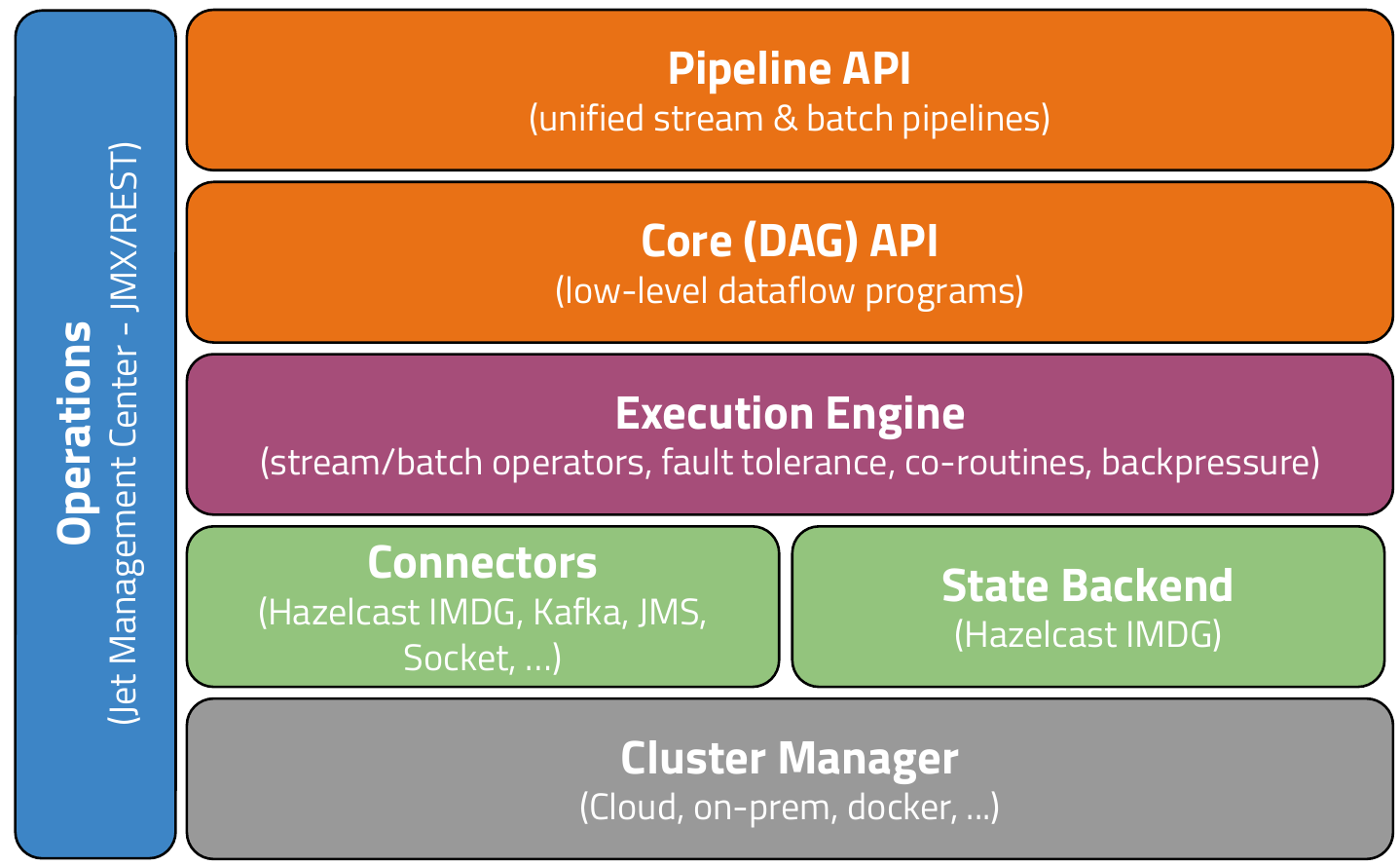}
\caption{Jet's Software Stack}
\label{fig:dag-distributed}
\vspace{-5mm}
\end{center}
\end{figure}

\section{Introduction}

Stream processing has completed more than twenty years of existence. Throughout the years, the database community has focused on defining semantics for streaming queries as well as system designs and optimizations. During the last seven years, stream processing is becoming mainstream: it is used for real-time equipment monitoring, dynamic car-trip pricing, credit card fraud detection, real-time harvesting of analytics, maintaining materialized views, and many other use cases and applications. 

\blfootnote{$^*$Work done while at Hazelcast.}

The needs of these new applications are starting to push existing streaming systems to their limits; each application may require a different system design. IoT applications and the banking industry, for example, have very different infrastructure and system design requirements. As a result of maturation, the stream processing landscape is now witnessing the well known ``one size does not fit all'' principle \cite{onesize}: in an attempt to cover different application and deployment requirements, multiple streaming systems are being developed, each specializing in its own niche. For instance, Apache Kafka \cite{kreps2011kafka} focuses on storing and integrating streams, Apache Flink \cite{CarboneKE15} enables out-of-order processing~\cite{li2008out} with a very general programming model, and Heron~\cite{KulkarniBF15} retains backwards compatibility with Storm, and focuses on flexible container-based execution.

Most of these systems were designed with specific use cases in mind, while their ability to adhere to strong SLAs has been an afterthought: the way that these systems are architected does not focus on low latency, embedded deployments. When tested in practice, most existing scale-out stream processing systems fail to keep the 99.99th percentile latency at low levels (e.g. tens of milliseconds). In fact, even with resource over-provisioning, latency on the 99th percentile can easily reach seconds in state of the art stream processors~\cite{karimov2018benchmarking}. We believe that this is the result of the design decisions behind these systems: in order to provide easy programming models and out-of-order processing \cite{akidau2015dataflow} as well as flexibility in deploying streaming jobs in the cloud, existing streaming systems abstract away from the hardware and rely too much on the host language amenities such as the garbage collection on the JVM.

At Hazelcast, around the end of 2015 we navigated the space of streaming systems and we concluded that none fulfilled our tight SLA requirements. In this paper we present Jet, a distributed streaming system that was designed with the latency at the 99.99th percentile as its primary focus. In the next paragraphs, we describe the motivation and discuss why we developed Jet. 

\para{Hazelcast's In-memory Data Grid}
Hazelcast IMDG is an open-source\footnote{https://github.com/hazelcast/hazelcast}, distributed, in-memory object store supporting a wide variety of data structures and interfaces inspired by the standard Java libraries. IMDG's data structures include \texttt{Map}, \texttt{Queue}, \texttt{Ringbuffer}, etc. For example, the \texttt{Map} interface, provides a distributed in-memory key-value store that Jet uses for storing its snapshots (\autoref{sec:snapshots}). Hazelcast IMDG's data structures are highly scalable and available: keys are stored in data partitions which are replicated across multiple machines in the cluster for fault resilience and parallelization. As IMDG members join and leave the cluster, partitions are re-balanced automatically. Finally, IMDG's data structures have AP behavior \cite{cap}: IMDG's data structures maintain linearizability when there's no network partition and prefer availability during them..

\para{From IMDG to Jet}
With time, we noticed that more and more of IMDG's users employed it for computation-heavy tasks. Although IMDG allows for very fast propagation of individual updates to replicas of our data structures, our users wanted to perform more \textit{global} operations: their tasks included joins among multiple distributed, and replicated data structures as well as aggregates that had to be updated whenever updates happened to different distributed IMDG data structures. To allow for such global queries, our team at Hazelcast first attempted to build a MapReduce-like prototype, only to realise after some time, that the requirements of our clients for very low latency were not covered by a batch-oriented architecture. 

For this reason, that batch prototype was quickly discarded in favor of a full blown streaming dataflow  processing engine that makes use of IMDG - both as a means to retrieve data but also to store data. To eliminate latency issues we decided that the partitioning of IMDG would have to \textit{align} with the partitioning of our execution engine, to avoid costly and slow data re-partitioning. To this end, we built Jet\footnote{http://jet-start.sh}, an open-source\footnote{https://github.com/hazelcast/hazelcast-jet} parallel dataflow processor that can execute custom business logic using data from the entire IMDG cluster.

\para{Main Design Decisions} 
Jet was initially designed for continuous, low-latency computations over streams of changes, data structure updates, etc. The system design first focused on making operators as lightweight and performant as possible for scale-up computations on a single node \cite{mcsherry2015scalability}. At the same time, our primary development platform is the Java Virtual Machine (JVM) which is not ideal for low-latency systems. To this end, we had to make careful design decisions in order to: $i)$~optimize the use of threads, $ii)$ minimize the effects of the JVM and its garbage collector to the execution engine, and $iii)$~minimize the system's memory footprint. To be amenable to embedded deployments in constrained environments, our team has been working hard to make sure Jet does not have any external dependencies; Jet can be deployed as a single Jar file. 

A single node running Jet has been shown to aggregate 10 million events per second with a 99.99th percentile latency below 10 milliseconds\footnote{https://jet-start.sh/blog/2020/08/05/gc-tuning-for-jet}. Jet can be deployed with minimal memory and computational requirements, while it can support hundreds of concurrent jobs within the same JVM (\autoref{sec:experiments}). In the rest of the paper we outline the design of Jet, its main contributions in the systems-engineering space as well as the lessons learned from building it. 

This paper is structured as follows.
Next, Section~\ref{sec:system-overview} provides an overview of Jet's core components: its programming model, execution engine, and state backend.
Section~\ref{sec:execution-model} elaborates Jet's execution model followed by Section~\ref{sec:state}, which presents Hazelcast's in-memory grid and how it supports state management, fault tolerance, and reconfiguration.
Section~\ref{sec:low-latency} describes how Jet's components come together to achieve single-digit millisecond latency at the 99.99th percentile. Section~\ref{sec:use-cases} narrates some of the interesting use cases that Jet supports and Section~\ref{sec:experiments} complements that description with a thorough experiment including performance microbenchmarks, fault tolerance measurements, and multitenancy scenarios on queries of the NEXMark benchmark.
Section~\ref{sec:related-work} presents related work and how Jet resembles but also differs from existing systems.
Notably, Section~\ref{sec:future-jet} introduces our vision for Jet as a serverless platform for running event-driven applications, such as microservices and stateful applications.
Finally, Section~\ref{sec:conclusions} concludes this paper.
















\section{System Overview}
\label{sec:system-overview}

Jet adopts the standard streaming data flow model, in which computations form a directed acyclic graph of operators (\textit{vertices}) that apply transformations on data streams (\textit{edges}). Jet can operate equally well as an embedded stream processor with a very low footprint, as well as a scale-out system that spans many nodes. 

Jet offers three main entry points for users: $i)$~the Jet Management Center -- a web UI and REST API from where users can manage and monitor Jet jobs, $ii)$~the \texttt{Pipeline} API -- a high-level API that can be used to create streaming pipelines and $iii)$  the \texttt{Core} API --  a lower-level API that can be used to specify low-level execution strategies. Both APIs are used to create distributed, stateful, and fault-tolerant dataflow programs. We detail those below.

\subsection{The \texttt{Pipeline} API}
\label{sec:pipeline-api}
The \texttt{Pipeline} API is designed to be the primary API that Jet users face. It offers transformations such as \texttt{map}, \texttt{filter}, \texttt{aggregate}, etc. that users can use to build complex data flows. It very much resembles Java streams in that, it is a fluent API and it is type-safe: the operators' input and output types are checked at compile-time. You can control some of the lower-level aspects at this level, too, like operator parallelism and traffic rebalancing. The example in \autoref{lst:wordcount}, shows the Java API of the typical ``Word Count'' program. 

\begin{figure}[t]
\begin{lstlisting}[style=javalang, numbers=none, xleftmargin=0in, label={lst:wordcount}, caption={Word count in Jet's Pipeline abstraction.}]
Pipeline p = Pipeline.create();

p.readFrom(lineSource())
 .flatMap(e -> traverseArray(e.
             toLowerCase().split("\\W+")))
 .filter(word -> !word.isEmpty())
 .groupingKey(wholeItem())
 .aggregate(AggregateOperations.counting())
 .writeTo(someSink());
\end{lstlisting}
\end{figure}

\para{Unifying Batch \& Streaming in a single API} The \texttt{Pipeline} API models computations as \textit{stages}; a stage represents what is known in different dataflow systems as an \textit{operator}. Stages can be either \textit{streaming} or \textit{batch} and they can be mixed and matched to build hybrid stream-batch processing data flows. The difference between the two is rather simple: streaming stages assume that their inputs are infinite, while batch stages assume that their inputs are finite.
For instance, as listed in \autoref{lst:hybrid}, one can build a pipeline that performs a hashjoin between a batch ``build side'' stage and a streaming ``probing side'' stage. The batch side will pull all the inputs of the batch stage performing complex batch filters, aggregates, etc. when the pipeline initializes, and then the stream will simply probe the hashtable for each incoming event on the streaming side.

In short, the \texttt{Pipeline} API serves as syntactic sugar that simplifies building scalable Jet workflows: pipelines are actually translated to parallel, distributed DAGs of operators at the \texttt{Core} API which we detail in the following text.

\subsection{The \texttt{Core} API}
\label{sec:core-api}
The \texttt{Core} API fully exposes the capabilities of the core execution engine. It can be used to build DAGs of operators with a finer level of control than the \texttt{Pipeline} API, but at the cost of increased complexity and the lack of almost all sanity checks, including type safety. Even though it is possible, this API is not intended to create DAGs by hand. Instead, it offers the infrastructure to build high-level DSLs and APIs that describe dataflows. It serves as a form of intermediate representation and requires familiarity with concepts like partitioning schemes, distributed vs. local edges, watermarks, etc.

The \texttt{Core} API can be used to fine-tune the execution properties of streaming DAGs. For instance, one can define the size of the queue between two vertices, a custom partitioning strategy of a distributed edge, or create custom code on how to react to an incoming watermark.

\begin{figure}[t]
\begin{lstlisting}[style=javalang, numbers=none, xleftmargin=0in, label={lst:hybrid}, caption={Hybrid Batch \& Streaming Program.}]
Pipeline p = Pipeline.create();
BatchStage<Person> persons = p.readFrom("persons");
BatchStage<Tuple2<Integer, Long>> countByAge = persons
  .groupingKey(Person::age)
  .aggregate(counting());
StreamStage<Order> orders = p.readFrom(kafka(...));
StreamStage<Entry<Order, Long>> ordersWithAgeCounts = orders
  .hashJoin(countByAge, joiningMapEntries(Order::ageOfBuyer))
  .writeTo(someSink());
\end{lstlisting}
\end{figure}

\subsection{Execution Engine}
\label{sec:execution-engine}

The execution engine contains implementations of very efficient operators for partitioning, window aggregation \cite{tangwongsan2017low, traub2019efficient}, joins, as well as the base source and sink operators, etc. Jet implements a unique way of sharing the computational resources, called tasklets (\autoref{sec:tasklets}), a form of co-routines that are cooperatively scheduled in a fixed pool of worker threads as shown in \autoref{fig:cooperative-threads}. The execution engine also performs state management (\autoref{sec:state}) and fault tolerance \autoref{sec:snapshots} that we present later in this paper.

\subsection{State Backend: Hazelcast IMDG}
\label{sec:imdg}
Unlike most streaming systems that store their snapshots in stable object storage like Amazon's S3, Jet uses IMDG for storing snapshots in a partitioned and replicated manner. The partitioning of the state stored in IMDG matches the partitioning of Jet. As a result IMDG's replicated partitions allow Jet to $i)$ retain its low latency by recovering very quickly from a node holding a state replica in case of failure, and $ii)$ to be able to elastically scale-out when the workload increases. We detail these in \autoref{sec:snapshots}.

\section{Execution Model}
\label{sec:execution-model}

Jet's execution model involves the deployment of stateful dataflow graphs that we explain in Section~\ref{sec:dataflow-graphs} and the execution of computations (Section~\ref{sec:tasklets}).
Section~\ref{sec:backpressure} clarifies how Jet handles backpressure.

\subsection{Deploying Stateful Dataflow Graphs}
\label{sec:dataflow-graphs}

\begin{figure}[b]
\begin{center}
\includegraphics[width=\columnwidth]{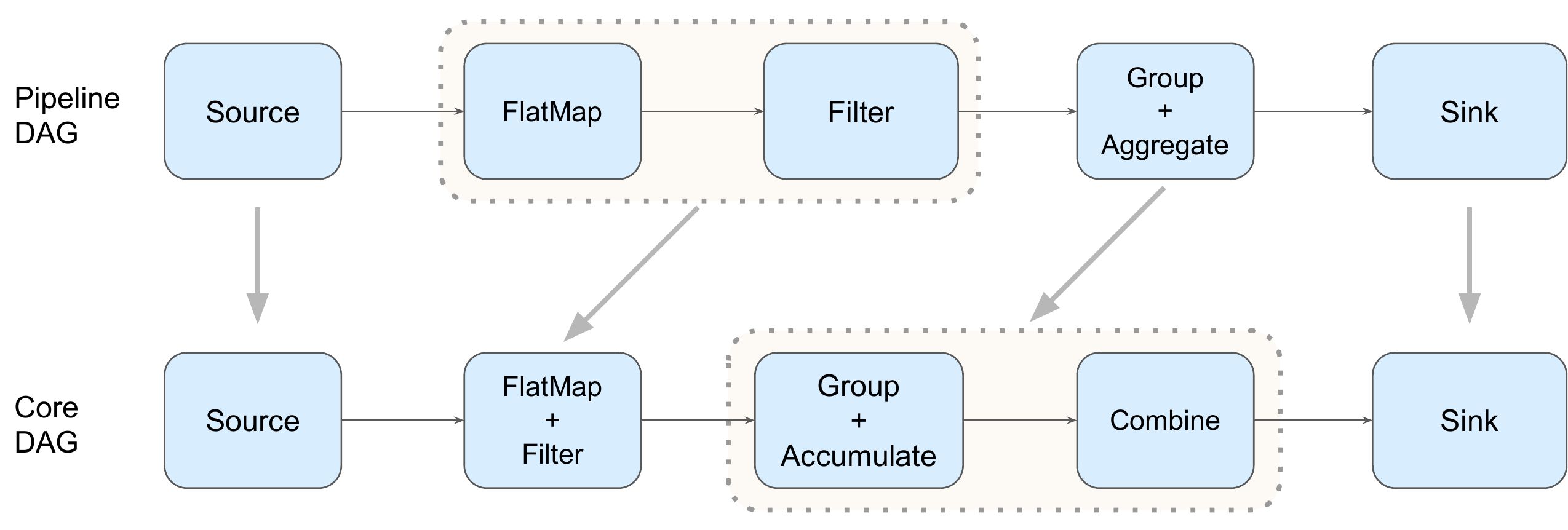}
\caption{Logical \texttt{Pipeline} of a Jet job (top) and a \texttt{Core} DAG in which Jet applied operator fusion (bottom).}
\label{fig:dag-fusion}
\end{center}
\end{figure}



The logical dataflow graph from the \texttt{Core} API is parallelized as in most streaming systems \cite{ArmbrustDT18, CarboneKE15}, i.e., massively parallel operators like maps and filters simply receive arbitrary parts of the stream, while \textit{keyed} operators such as joins, and aggregates execute in a partitioned manner. However, Jet does not follow the typical operator-per-core model. Instead, it deploys the \textit{complete} dataflow graph on every available CPU core as seen in \autoref{fig:dag-distributed}. Since the source and sink connectors depend on 3rd-party APIs, and some of these rely on blocking calls that cannot be made to work cooperatively, Jet must start dedicated threads for them. After empirical experimentation we decided to deploy at most two such operators per node. For instance, \autoref{fig:dag-distributed} depicts one source and one sink on each node. 

Deploying the complete dataflow graph on every single CPU core has several advantages with respect to performance. The most important is that it allows to keep data exchange local to the machine as much as possible, avoiding expensive data transfers across the network. Second it allows for a coroutine-based execution that we discuss in the next section. 

Jet further minimizes communication costs between operators in three ways. First, it fuses (a.k.a. operator chaining) consecutive stateless operators \cite{hirzel2014catalog}. Second, it keeps data exchange local whenever possible, even on partitioned data exchanges, where it applies a two-stage approach (local partial results followed by global combining). Finally, when an operator needs to pass events to a remote operator, Jet deploys an exchange operator \cite{graefe1990encapsulation} on the sender and receiver sides that takes over the data partitioning and exchange. In summary, Jet leverages both pipeline parallelism and data parallelism within each node but also across nodes. Together with coroutine-based execution (\autoref{sec:execution-model}) Jet manages extremely low-latency stream processing.


\begin{figure}[t]
\begin{center}
\includegraphics[width=\columnwidth]{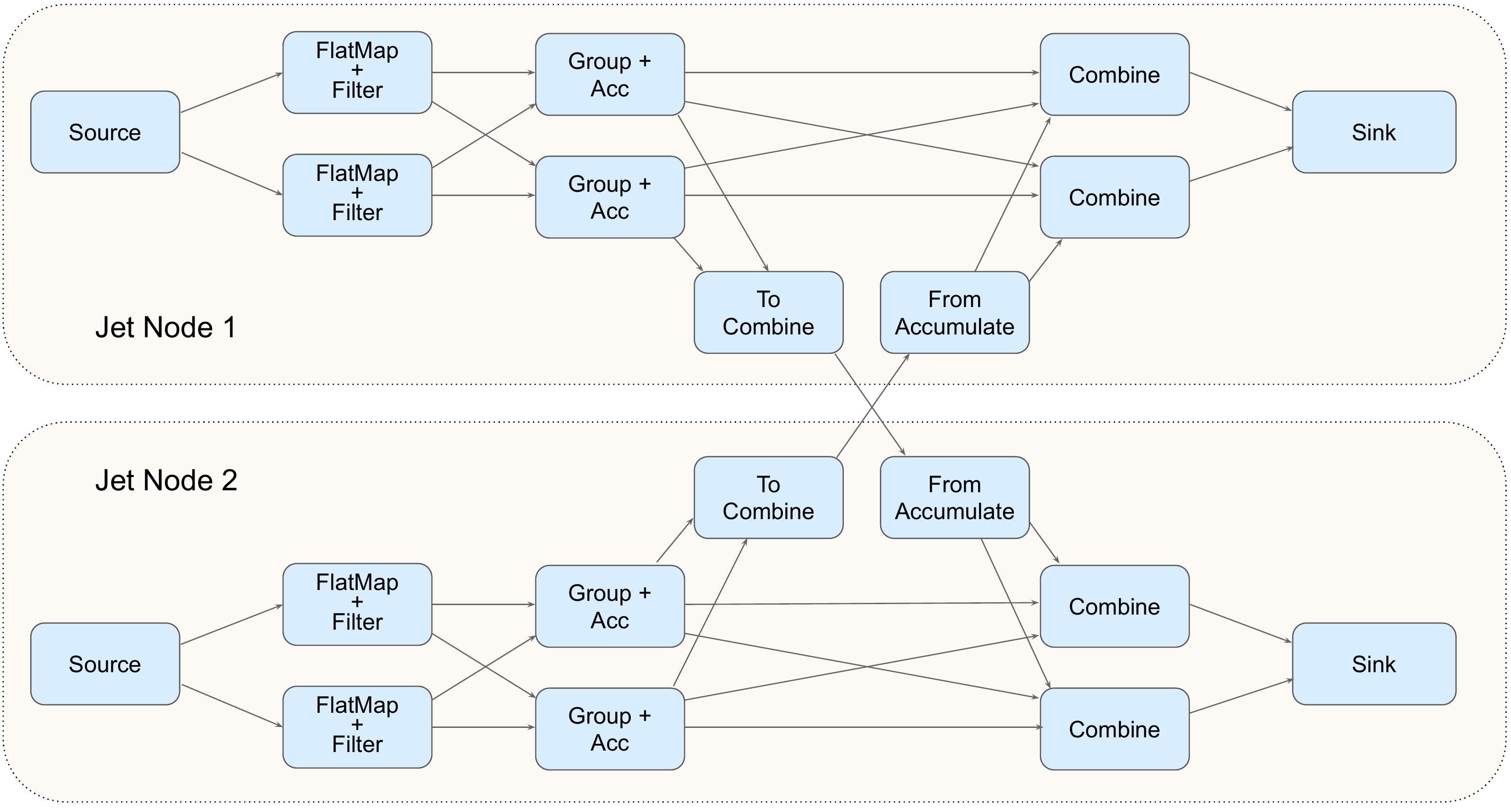}
\caption{Dataflow deployed with local parallelism of 2.}
\label{fig:dag-distributed}
\end{center}
\end{figure}

\subsection{Tasklets \& Cooperative Threads}
\label{sec:tasklets}

\begin{figure}[t]
\begin{center}
\includegraphics[width=.76\columnwidth]{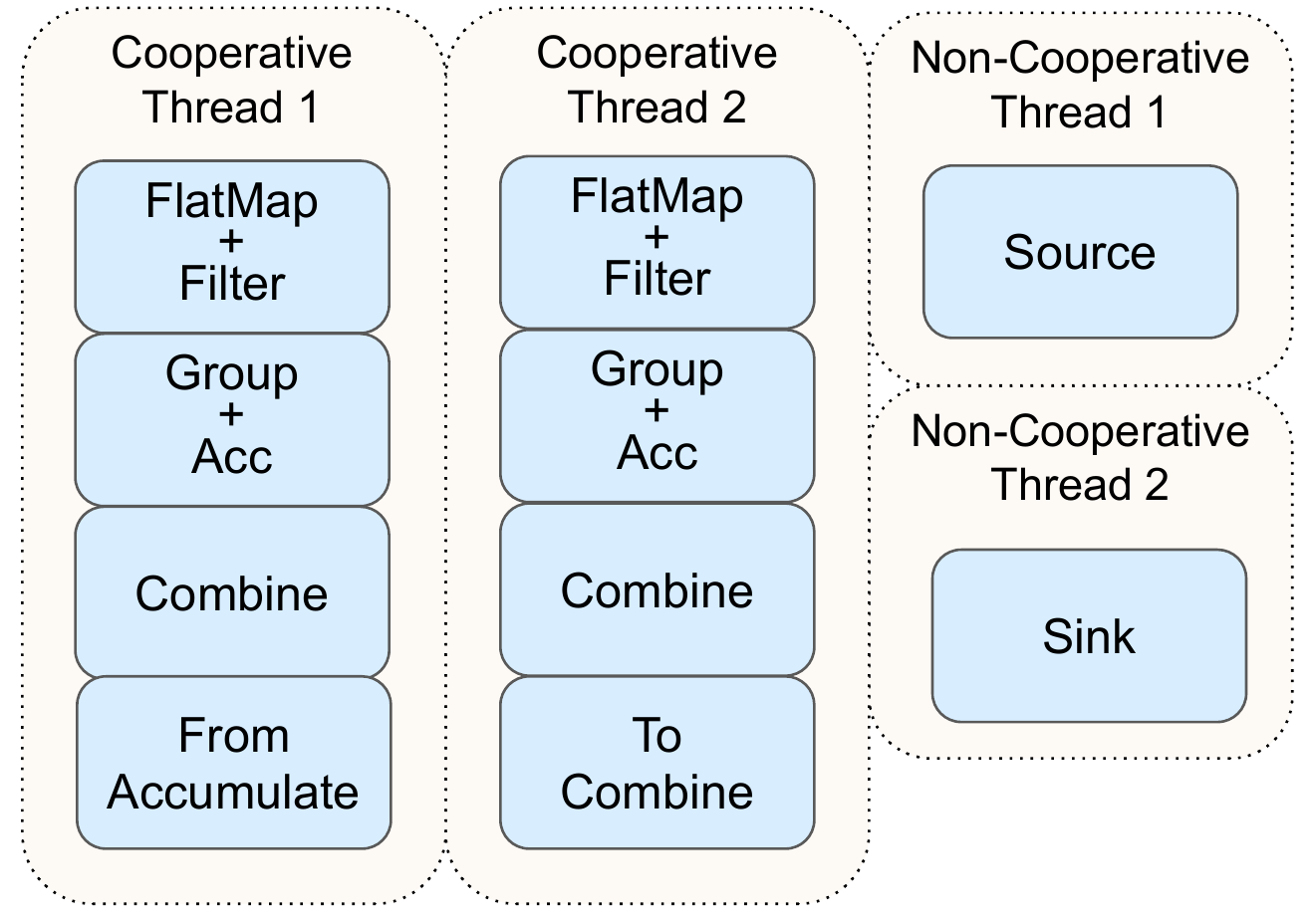}
\caption{Tasklet scheduling in Jet}
\label{fig:cooperative-threads}
\end{center}
\end{figure}

Jet's execution model is based on \textit{tasklets}, very small computational units that share an execution thread by taking turns in using a CPU core. As seen in \autoref{fig:cooperative-threads}, Jet deploys as many JVM threads as there are CPU cores. 
Instead of allocating one thread for each parallel tasklet, a thread takes over the execution of a number of tasklets. On each thread, Jet runs a loop that executes its tasklets in a round-robin fashion.

Tasklets, are a concept similar to \textit{coroutines} \cite{kahn1976coroutines}: they are small computation units that can be suspended and resumed at the programming language level.
More specifically, a tasklet voluntarily yields control to the Jet framework after executing for a very short period of time, typically under 1 millisecond.
Then Jet can schedule another tasklet to run on the same thread. 
This design achieves high CPU utilization by avoiding the costly operating system context switches and allowing execution threads to remain on the same CPU core for longer time periods, thereby preserving the CPU cache lines. A tasklet is guaranteed to perform meaningful work (e.g., perform an aggregate or a join of two events) in each very short execution period because it has no dependencies to the outside world. Jet owes much of its performance capacity to tasklets.

\para{Cooperative Threads} A collection of tasklets collectively occupy a \textit{cooperative} thread \cite{abadi2009model}. Tasklets start execution when their turn comes, and stop execution on their own accord.
Tasklets allow Jet to sustain high load and extremely high numbers of tasks on modest resources:  Jet can host tens of thousands of tasklets on a single execution thread, allowing for very large numbers of operators in a single node. For the same reason, Jet lends itself really well to multi-tenancy: one can run thousands of concurrent jobs on the same set of resources with very little overhead. 
This is because when a tasklet has no work to do -- typically because there is no input available -- it backs off from the thread and gives its place to another tasklet, without the need for context switching.

Although very performant, the tasklets execution model entails an important restriction: a tasklet should never block, because that would jeopardize the execution progress of all tasklets sharing the execution thread. Although Jet could use a cooperative scheduler, in practice we have seen that simply iterating over all tasklets repeatedly works pretty well. 
However, Jet does schedule \textit{blocking} computations on separate, \textit{non-cooperative}, dedicated threads so that they do not interfere with the execution of tasklets.
Non-cooperative computations are forced to frequently return control to Jet -- at least every second.
Jet needs to gain control of all execution threads in order to take state snapshots for fault tolerance strategy (\autoref{sec:state}).

\para{Jet Processors}
A \textit{processor}, implements the custom logic (e.g., the UDF, or the operator logic such as joining two tuples) of a given DAG vertex in a Jet dataflow graph.
Each processor includes an inbox of input records to be processed and an outbox of output records to be dispatched downstream.
A tasklet manages the processor's inbox and outbox, its state, and its inbound and outbound queues. A tasklet repeatedly calls the processor's \texttt{process} method until the processor has consumed all the input records from the inbox, and then the tasklet refills the processor's inbox with more input (possibly from a different input edge).
For each output record in the outbox, Jet, according to the record's key, finds the record's downstream destination by computing the partition ID, and forwards that record to the respective downstream vertex.

Tasklets within the same node exchange data through shared-memory, single-producer-single-consumer queues that use wait-free algorithms.
Specifically, a tasklet produces output to a queue that connects it with a specific consumer tasklet downstream.
The consumer tasklet can retrieve its available input from the queue to fill in its processor's inbox by pulling the records from these queues.
The producer tasklet stops producing records if the consumer tasklet's queue fills up.
In essence, a queue instance serves the connection between a pair of tasklets, and on the receiver and producer sides Jet uses wait-free algorithms for predictable and performant data exchange.

\subsection{Handling Backpressure}
\label{sec:backpressure}

Jet employs a backpressure mechanism to limit the amount of items a source vertex sends to the destination/consumer when that consumer throttles.  If this happens, streaming systems require a backpressure mechanism to signal back to the source to moderate its operation so that the whole pipeline stays in balance and operates at the speed of the slowest vertex. Local backpressure between tasklets inside the same Jet node is simple: Jet uses bounded queues and tasklets to back off as soon as all their output queues are full.

Backpressure is trickier over a network link. Jet uses a design very similar to the TCP/IP adaptive receive window: the producer must wait for an acknowledgment from the consumer specifying how many data items the producer can send. After processing item \texttt{n}, the receiver sends a message that the sender can send up to item \texttt{n + receive\_window}. The consumer sends the acknowledgment message every 100ms: as long as the \texttt{receive\_window} is large enough to hold the number of events processed within 100 milliseconds (plus network link latency), the receiver will always have data ready to be processed.  Jet calculates the size of the \texttt{receive\_window} based on the rate of event processing in a given tasklet and adaptively shrinks and expands the \texttt{receive\_window} as the flow changes. In stable state the \texttt{receive\_window} contains roughly 300 milliseconds' worth of data.

\section{State Management}
\label{sec:state}

Jet stores state exclusively in memory, through the use of IMDG's \texttt{IMap} data structure. This section describes how Jet organizes computation state across a cluster (Section~\ref{sec:state-org}) and then presents the function of the in-memory grid and how it supports fault tolerance and reconfiguration (Section~\ref{sec:imdg-state}). 
Then, Sections~\ref{sec:elasticity} and~\ref{sec:snapshots} describe Jet's overall approach to elasticity and fault tolerance.
Finally, Section~\ref{sec:assumptions} explains how Jet interacts with external source and sink systems to provide various levels of processing guarantees.

\subsection{State Organization}
\label{sec:state-org}

\begin{figure}[t]
\begin{center}
\includegraphics[width=\columnwidth]{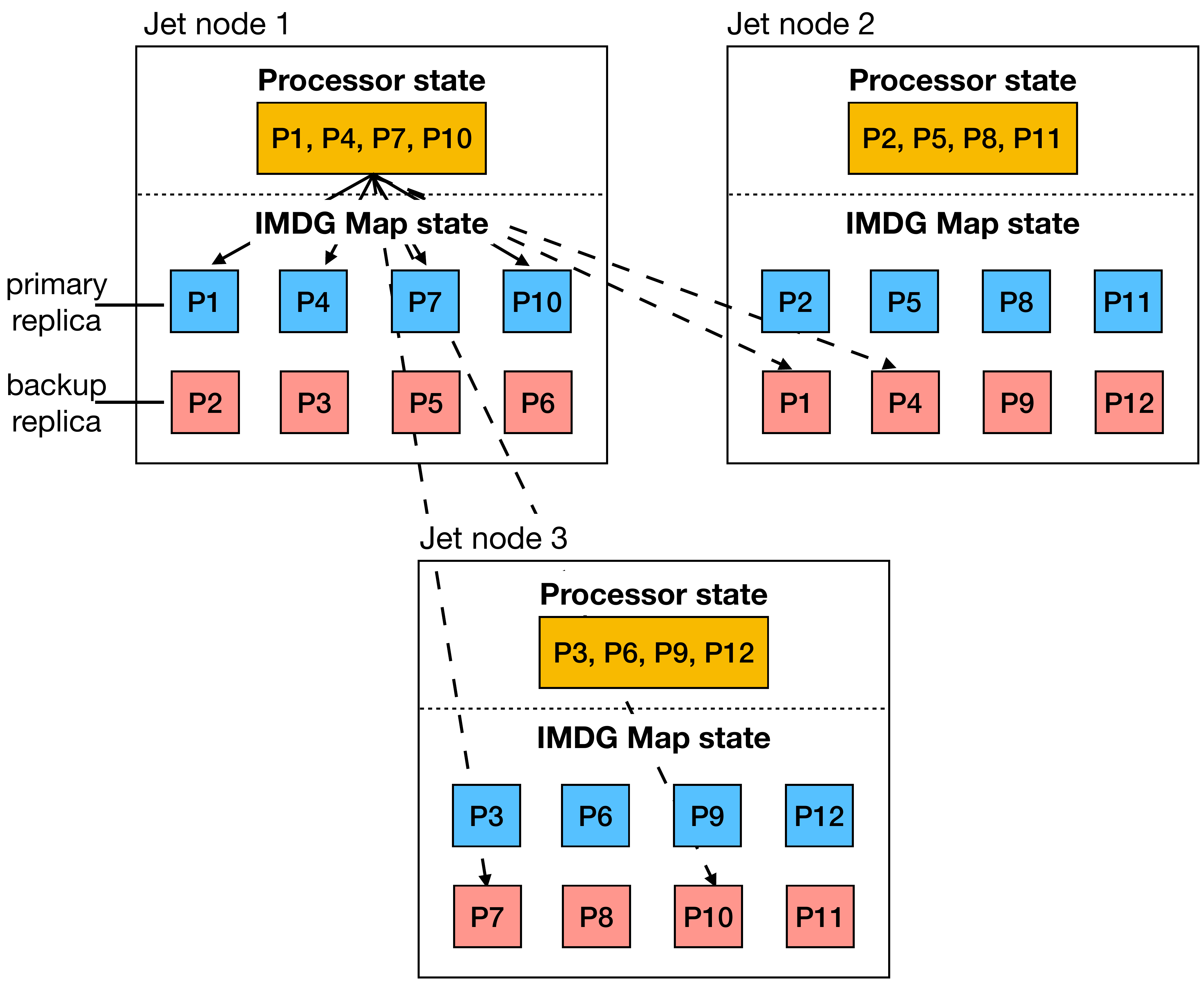}
\caption{Snapshot partitioning and replication in IMDG.}
\label{fig:imdg-snapshots}
\end{center}
\end{figure}

State in Jet is a collection of key-value pairs, held in memory (Figure~\ref{fig:imdg-snapshots}).
Jet uses IMDG's \texttt{IMap} data structure to partition the key space into disjoint partitions that are allocated to parallel instances of a stateful vertex. Partitioned distributed edges connect keyed stateful computations such as aggregations, with all downstream vertices. Stateful computations in Jet are supported by equally distributing disjoint partitions of the state to a stateful vertex's instances across all cluster nodes as shown in Figure~\ref{fig:imdg-snapshots}.
State is partitioned by a record key such that each vertex instance stores the state corresponding to a specific key space.
Thus, each record is shipped through the edge that links with the vertex holding the partition that matches the record's key.

\begin{figure}[t]
\begin{center}
\includegraphics[width=\columnwidth]{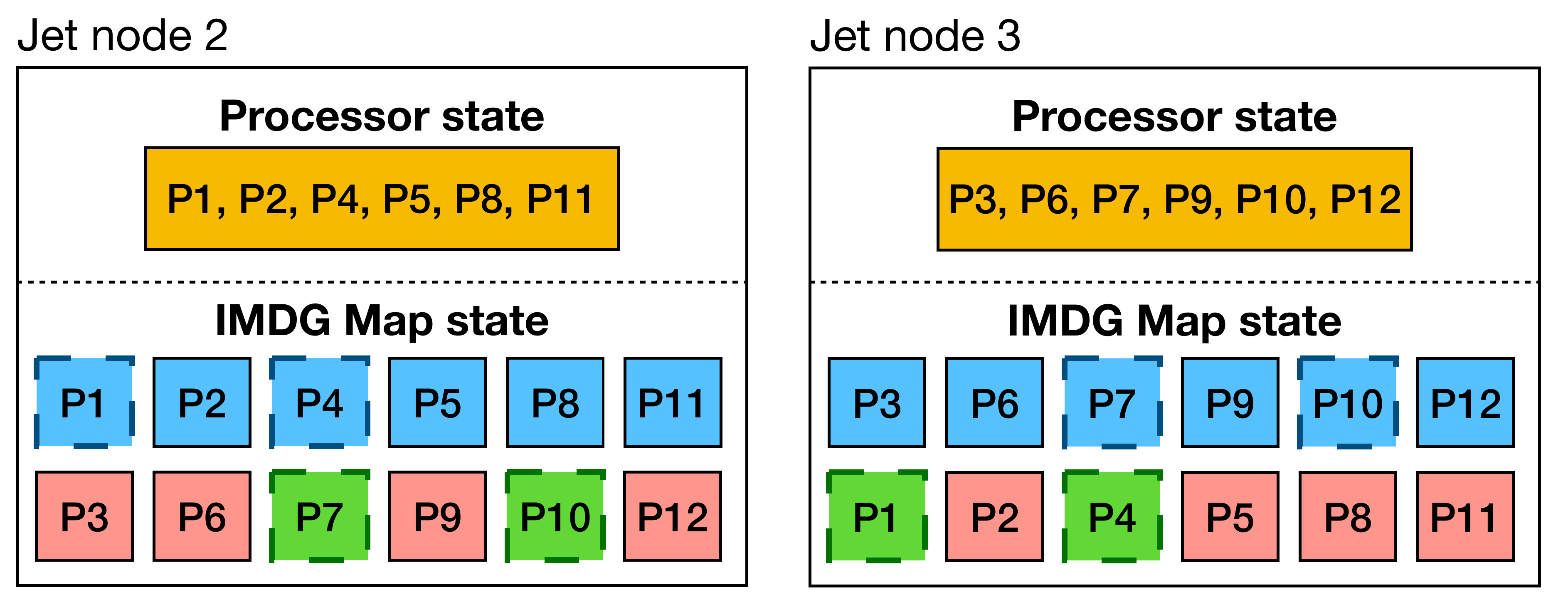}
\caption{Snapshot re-balancing on fault recovery.}
\label{fig:imdg-recovery}
\end{center}
\end{figure}

\subsection{In-Memory Data Grid (IMDG)}
\label{sec:imdg-state}

One key feature that distinguishes Jet from other modern scale-out streaming systems is that, although fault-tolerant and highly available, Jet has zero dependency on disk storage.
Instead, it uses the in-memory data grid, for state management and disaster recovery.
The IMDG implements a concurrent, distributed, observable and queryable map data structure.
Figure~\ref{fig:imdg-snapshots} depicts how IMDG supports a distributed dataflow graph.

A Jet processor operates directly on a local \texttt{Map} data structure (e.g., \texttt{HashMap}), state snapshots of which are stored in \texttt{IMap}. To optimize for locality, the partitioning of a Jet vertex matches the partitioning of the \texttt{IMap} data structure running on the same node. For each IMap partition, IMDG stores $i)$~a \textit{primary replica} on the same node and $ii)$~one or more backup replicas on other nodes, according to the IMDG cluster configuration. The backup replicas of that state snapshot are distributed to other member nodes of the cluster for high availability. Figure~\ref{fig:imdg-snapshots} highlights how state objects of the processor instance in node 1 are stored in the local node's primary replicas matching the partitioning scheme used for both processing and state snapshot storage. Additionally, backup replicas of these state objects are sent to the 2nd and 3rd node.

When a member node in a Jet cluster fails, e.g. node 1 in Figure~\ref{fig:imdg-snapshots}, the stored replicas of that node are lost together with the state of the Jet processor instances.
In this case, IMDG has stored backup replicas of node 1, in nodes 2 and 3.
To recover node 1, Jet instructs node 2 to promote partitions P1 and P4 from backup to primary as Figure~\ref{fig:imdg-recovery} shows. Node 3 performs the same for partitions P7 and P10.
Nodes 1 and 2 produce backup replicas for each other's newly added primary replicas.
As a final step, Jet populates processor state from the corresponding primary replicas.

\subsection{Elasticity and Reconfiguration}
\label{sec:elasticity}

Beyond fault tolerance, IMDG is the key enabler for Jet's elasticity and reconfiguration functionality.
When a new node enters a Jet cluster, Jet assigns replicas to that node in order to rebalance the partitions.
During the rebalancing phase, Jet minimizes data migration between the nodes employing consistent hashing \cite{stoica2001chord}.
For the job depicted in Figure~\ref{fig:imdg-snapshots}, the new node is assigned the primary replica of P1 from node 1 and the backup replica of P2 again from node 1.
Following the rebalancing phase, the job is restarted with processor states initialized from the local primary replicas of the latest state snapshot.

\subsection{Fault Tolerance and Processing Guarantees}
\label{sec:snapshots}

Jet's approach to fault tolerance is based on the seminal work on distributed snapshots by Chandy and Lamport\cite{chandy1985distributed} that has been applied to stream processing~\cite{CarboneKE15, CarboneEF17} and adopted by many streaming systems~\cite{CarboneKE15, SilvaZD16, ChandramouliGB15}.
The approach involves periodic checkpoints that produce a consistent global snapshot of the system's state.

At regular intervals, Jet instructs source vertices to take a state snapshot.
Then, all processors belonging to source vertices save their state, emit a checkpoint barrier to the downstream processors through the data flow, and resume processing.
When the checkpoint barrier reaches the sink processors, all processors have saved their state to stable storage and the state snapshot is complete. To recover from a failure, Jet will stop processing in all nodes and vertices, reload the latest state snapshots from IMDG recorded at the latest checkpoint, spawn a new instance to substitute the one that failed, and ask the input sources to replay the input data following the latest checkpoint.

The described protocol can offer \textit{exactly-once} processing guarantees \cite{CarboneEF17}, i.e., each input will leave its effects on the system's state exactly-once despite failures. Exactly-once consistency entails that no processing is performed while a checkpoint takes place. In the case of multiple input channels, once a checkpoint barrier arrives at one input channel, that channel needs to block and wait for all checkpoint barriers to arrive at the rest of the input channels of that vertex. For \textit{at least-once} processing guarantees channels do not need to block, decreasing latency. Jet offers both levels of consistency as part of its configuration.

\subsection{Assumptions and External Systems}
\label{sec:assumptions}

Jet and most streaming systems~\cite{ArmbrustDT18, CarboneKE15, NoghabiPP17, SilvaZD16} rely on two assumptions for ensuring \textit{exactly-once} guarantees.
In essence, these assumptions complement the system's fault tolerance approach.
The assumptions regard the capabilities of systems functioning as data sources and data sinks to Jet.

Jet requires that source systems, which provide the input data, are \textit{replayable} or \textit{acknowledging}.
A replayable source can replay the input data it produces from a specific offset.
Jet needs this functionality when it is recovering from a failure in order to provide exactly-once processing guarantees as we described in Section~\ref{sec:snapshots}.
In fact, Jet's source vertices take part in the checkpoints to store the source offset at which a checkpoint is taken so that a potential replay in case of a failure is precise.

If a source system is not replayable, but accepts acknowledgements that the data it stores can be safely deleted, then Jet can provide \textit{exactly-once delivery} guarantee by acknowledging items only after they are processed by the entire pipeline and a successful snapshot has been taken. Since completing the snapshot and sending the acknowledgements isn't atomic, it can happen that the job fails before Jet acknowledges all the items. Because of this, Jet uses record IDs to deduplicate messages when the remote system re-sends unacknowledged messages after a recovery that had already been received by Jet prior to the failure. The record IDs are stored in a global state snapshot.

The exactly-once delivery guarantee can also be achieved on the output side of Jet if a sink system, which provides Jet's output to a consuming application, functions as a \textit{transactional sink}~\cite{CarboneEF17} or supports \textit{idempotent writes}.
A transactional sink withholds output and only makes it available to the outside world when a checkpoint is complete.
In essence, it performs a two-phase commit on the output received by Jet.
The commit-prepare phase executes when a checkpoint begins, with the second phase commit happening after the checkpoint is complete.
At this point, the state corresponding to the output has been persisted.

Alternatively, the exactly-once delivery guarantee can be achieved if the sink system supports idempotent writes.
Idempotent writes have the exact same effect irrespective of the number of times they are applied.
Idempotent writes obviate the need for deduplication.

\subsection{Fault Tolerance via Active Replication} 
\label{sec:ft-tradeoffs}

From our experience with clients and Jet users we have noticed a very interesting trade-off: by implementing an algorithm for making snapshots in streaming computations such as the seminal Chandy-Lamport algorithm or one of its variants, the system has to pay the price of synchronizing barriers and blocking the streaming pipelines - especially to ensure exactly-once processing. During the Jet development journey we have made an important design decision: instead of focusing on creating low-latency snapshots, we opted for optimizing our system towards the efficient use of the available resources. In our experiments (\autoref{sec:experiments}) we show that Jet can easily handle a throughput of 2M events per second per CPU core, with very low overhead in latency.
Thus, instead of running large deployments of a stream processor and requiring very efficient fault-tolerance mechanisms, we opted for enabling users to use less resources for a given workload, allowing them to run active-active deployments in which the job is executed twice (one active and one as active stand-by). The result is that in the absence of book-keeping and overhead for fault tolerance such a deployment can sustain failures, but it also performs extremely efficiently. In fact, this has proven quite a useful trade-off for our users and has simplified our system design. At the same time, we are considering alternative solutions like speculative replication \cite{DBLP:conf/sigmod/WangFAM20}.

\section{Millisecond Latency on the JVM}
\label{sec:low-latency}

The key competence of Jet is its capacity to perform ultra low-latency processing.
Four design decisions drive this achievement.
First, Jet schedules and executes threads at the programming language level leveraging the JVM in a way that resembles coroutines and green threads as we described in Section~\ref{sec:tasklets}.
Second, it minimizes the interference of concurrent garbage collection.
Third, it utilizes the distributed IMDG to provide scalable and reliable state management in memory.
Finally, Jet deploys a full dataflow graph instance to each member node of a cluster, thereby minimizing network connections for data transfers, which introduce latency.

\para{Optimized data path} Jet optimizes the data path from source to sink: in fact, virtually all components involved in the datapath have been (re)implemented multiple times in the course of the last four years.
Source vertices are local to each node and connect only with local vertices.
Input data is received by each vertex's concurrent queues, which supply the data to the tasklet that will process it.
Because tasklets do not depend on other computations, they can make progress in very short periods of time.
When a tasklet's window of execution expires, it voluntarily yields control back to Jet, which selects the next tasklet to execute on the same thread without depending on operating system level schedulers.
Periodically, Jet stores a state snapshot of the stateful vertex's state in the node's local IMDG space and replicates it to other nodes for high availability.
Consequently, the datapath remains local to a node as much as possible passing through fast execution stages, which operate on data kept exclusively in-memory.

\para{Garbage collection} A lot of attention and experimentation has been devoted to garbage collection in order to minimize its overhead to Jet's operation.
Notably, garbage collection is recognized as one of the hidden performance enemies of stream processing, especially in terms of latency, due to its interference with the scheduling of computations.

As we described in Jet's execution model in \autoref{sec:execution-model} Jet employs a thread pool with as many threads as a member node's CPUs, however it can be configured to use a few cores less. The rationale for this choice is to allow garbage collection to be concurrent to the execution of tasklets by functioning as a background service without interfering with the scheduling of computation threads on a CPU.
Dedicating a few threads to garbage collection and the remaining threads to computations optimizes the scheduling of threads to CPUs and ensures that each thread remains longer at a CPU.
This design allowed Jet to bring its latency to under 10ms without impacting the throughput, as we show in the experiments.

\section{Use Cases}
\label{sec:use-cases}

Hazelcast clients and open-source users deploy Jet in many different ways; many of which we found surprising. Here we report on a subset of them that we believe are closest to the interest of the SIGMOD community, and were not typical (e.g., windowed aggregates) uses of streaming technology. We gathered these use cases from our internal engineering teams, our mailing lists, and Jet's Slack channels.

\para{Real-time Rule Execution} A user from the banking industry aims at performing transactions. In order to check if a transaction is fraudulent or not, they need to first inspect multiple terabytes of data stored in a Hazelcast IMDG gathering ML features of a client and their previous transactions, before they can execute custom business logic, and decide whether a transaction is fraudulent. Because of the complexity of the process, Jet is assigned a maximum of 2ms for executing the complete set of tens of business rules. Similarly, another user from the banking industry combines complex business rules executing against the streaming operator state kept inside Jet operators. 

\para{Internet of Things} Since Jet is very lightweight and has very few dependencies, we have seen a set of interesting IoT deployments. Deployments in factories and vehicles perform typical metrics aggregation (speed, temperatures, etc.) used for monitoring purposes, with relaxed latency requirements (hundreds of ms latency). 

\para{View Maintenance} We observed that a lot of Jet users use it to maintain views: they subscribe to different IMDG objects and external databases and capture changes in data items in the form of Change Data Capture (CDC)\footnote{https://debezium.io}. Subsequently, they consume the CDC stream and build materialized views that are updated with every change that happens to the external data sources.

\para{Stateful AI} Finally, one of the most interesting use cases of Jet is its deployment as a stateful backend for a chatbot: the chatbot performs lookups for information that are related to humans' questions in a chatbox. The chatbot is deployed as an automaton where Jet operators are states and edges represent transitions. On each interaction with the human, the chatbot updates its state and responds to users. Our client scaled the chatbot to thousands of messages per second in a limited amount of computational resources.

\para{Oil Rig Drilling}
The operating costs of a drilling rig are very high and any downtime throughout the drilling process can have a significant effect on the rig operator’s bottom line. The rigs are equipped with a large number of sensors to detect small vibrations during the drilling process. Hazelcast Jet enables human operators to immediately act on the streaming data in near real-time to prevent costly, catastrophic failures in the drilling process.
It enables managing physical resources better through high-frequency feedback on a per-well basis, which amounts to reducing the drilling time by as much as 20 percent, from a typical 15 days to 12 days thereby saving up to millions of dollars.

The data input to Jet are generated from sensors.
At any time during rig operations, up to 70 channels of high-frequency data enter Jet at various frequencies.
Jet records and analyzes the data or events that occur, and then applies proprietary algorithms to make very fine-tuned adjustments to the drilling process. An example of this is the real-time adjustment of the revolutions per minute (RPM) of the drilling string and bit. A drill bit is a type of drill that must be operated at a very precise level to prevent equipment failure and costly delays to the drilling process.
Jet computes stateful aggregates over 10K messages/second maintaining latency under 10ms.
The workload resembles query 6 of NEXMark queries that we include in the experiments.

\para{Real-time Payments}
Hazelcast Jet is used as the backbone framework of an instant payment processing application. Within this application, Jet operates as a processing pipeline for each step of the payment process. The payment management application acts as an orchestrator which analyzes XML payment instructions and forwards them to the respective card’s issuing bank or card association for verification, and also carries out a series of anti-fraud measures against the transaction before settling the payment transaction.

The computation is a stateful complex workflow of 150 stages in total, which communicate with external systems, such as databases. Multiple Jet processing jobs (tens of jobs) are involved as pipeline components where Jet's DAG API is used for stream processing while its pipeline API is used for batch processing. Jet's API offers considerable flexibility in expressing sophisticated multi-stage computations and combining stream and batch processing stages.
The pipeline components rely on the distributed IMaps of Hazelcast IMDG for transaction ingestion and messaging through high-performance connectors that enable low-latency operations. Finally, the quick recovery mechanisms of the Jet cluster provides high availability to the instant payments application protecting it against failures.

The payments workload is CPU-bound with current throughput at the input side averaging at 10K transactions per second. It is served by a 3-node Jet cluster with 8GB RAM and an 8-core CPU per node. In this setting, throughput scales linearly with CPU cores until it reaches 65K operations per node. Then adding another node is necessary to keep the latency flat.
Jet keeps latency stable at 85ms on the input and 150 ms on the output depending on the number of workflow steps in a specific transaction flow. Each workflow updates in-memory business entities at a rate of 75 updates per transaction on average.

\section{Experiments}
\label{sec:experiments}

In this section, we present the experiments we performed with Jet.
First, we describe our experimental methodology (Section~\ref{sec:experiment-method}).
Second, we show experiment results with respect to throughput (Sections~\ref{sec:throughput}, \ref{sec:maxthroughput}) and latency (Sections~\ref{sec:latency-nexmark} and \ref{sec:micro}). Then, we measure the overhead of the fault tolerance approach to Jet's normal operation in terms of latency (Section~\ref{sec:latency-ft}).
Finally, we demonstrate Jet's capacity for running a big number of jobs on a small cluster with low latency (Section~\ref{sec:multi-tenancy}).


\subsection{Experimental Methodology}
\label{sec:experiment-method}

We evaluate Jet v4.3 in an Amazon Cloud cluster comprising 1, 5 and 10 c5.4xlarge machines with 16 vCPUs and 32 GB RAM each.
Jet runs on Oracle OpenJDK v15.0.1 with 12 cooperative threads per node and the G1 garbage collector is configured with a GC pause target of at most 5 milliseconds. It does most of the GC work concurrently.

\para{Query Workload} The workload used in the experiments consists of queries 1, 2, 5, and 8 of the NEXMark benchmark~\cite{tucker2002NEXMark} as described in the Apache Beam project.\footnote{ \url{ https://beam.apache.org/documentation/sdks/java/testing/nexmark/}}
The benchmark defines queries over people participating in auctions.

\begin{itemize}
\item Query 1 is a simple map converting an amount from one currency to another.
\item Query 2 is a simple filter that selects auctions based on auction numbers.
\item Query 3 is a join and filter that selects sellers in particular US states.
\item Query 4 is a join with custom window functions and aggregation that reports the average selling price for each auction category.
\item Query 5 is a sliding window with an aggregation to report which auctions have seen the most bids in a given window.
\item Query 6 illustrates a specialized combiner that reports the average selling price per seller for their last ten closed auctions.
\item Query 7 illustrates fanout using side input and selects the highest bids per period.
\item Query 8 is a join of the stream of (new) users with the stream of  auctions, reporting users that created an  auction in the last period.
\item Query 13 is a join between a stream of auctions and a bounded side-input.
\end{itemize}

\begin{figure}[t]
\begin{center}
\includegraphics[width=\columnwidth]{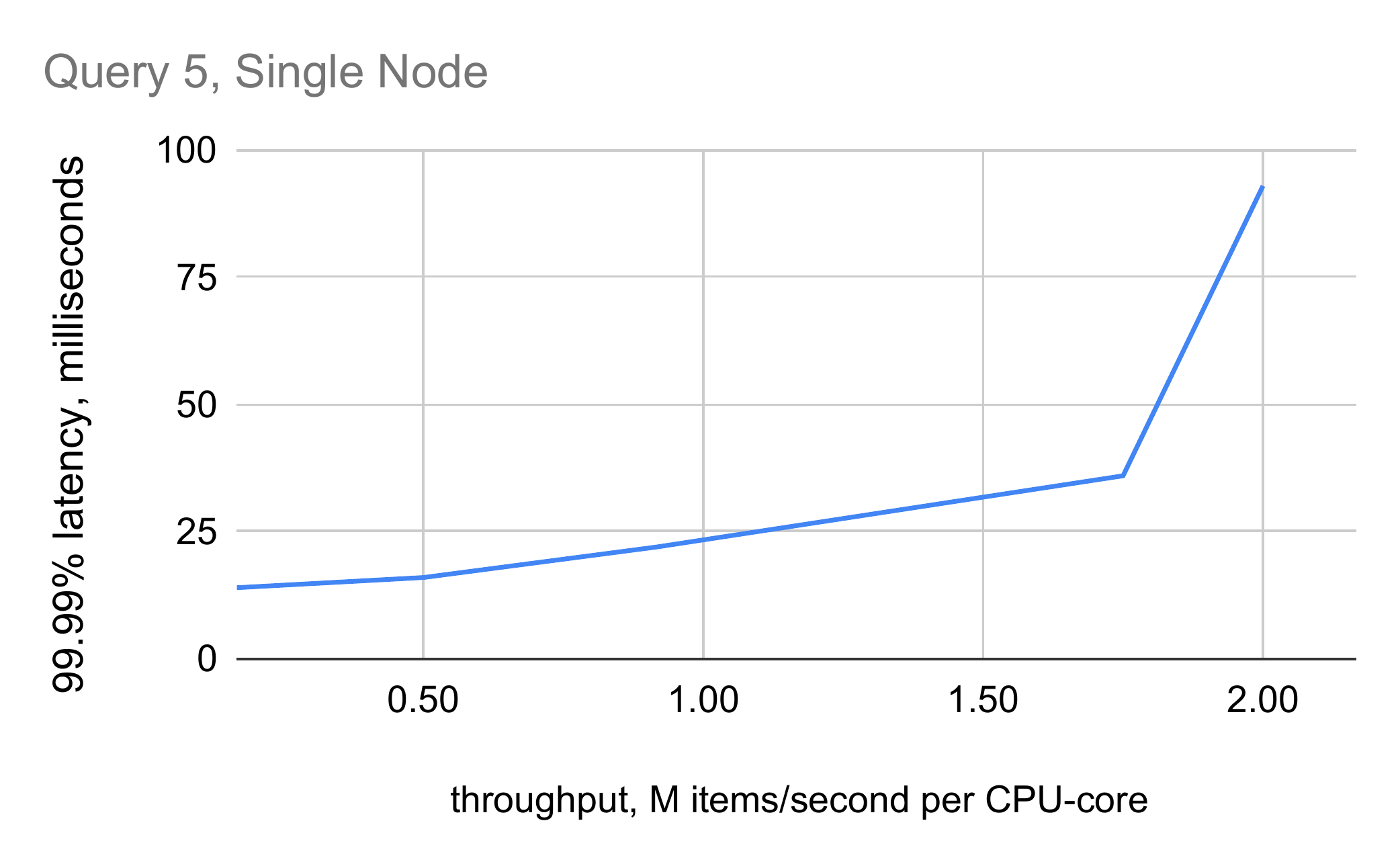}
\caption{Throughput per CPU-core vs. Latency for Q5 on a single node (12 CPU cores) with 10ms window slide.}
\vspace{-2mm}
\label{fig:throughput-per-core}
\vspace{-4mm}
\end{center}
\end{figure}

\para{Data Generator Configuration} We use the NEXMark data generator configured as follows. 
We define 10 thousand distinct keys that correspond to persons and auctions in the input dataset; we generate 1M records per second, by drawing keys randomly.

\para{Window size and slide}
In all windowed queries (Q5, 8, 13), unless stated otherwise, we use a 10-second sliding window join with a sliding step of 10ms. We choose to push Jet to its limits and show how it performs under pressure: triggering every 10ms is something that, to the best of our knowledge, no other scale-out stream processor can perform at the time of writing.

\para{Metrics \& Experiment Setup}
We first let the JVM warm up by letting the streaming job run for 20 seconds, then we run the measurements over a period of 240 seconds. With 100 sliding window results per second, that gives us 24,000 data points.
The primary metric of interest in our experiments is the processing \textit{latency}. Specifically, the latency clock starts for each event at its predetermined time of occurrence. At that point the source tasklet is allowed to emit it, but any delay in actually emitting it is already affecting the reported latency number. If the event's timestamp is such that it is the first event \emph{after} a given window's end time, it will trigger the aggregating stage to start emitting that window's result. The clock stops when Jet has started emitting the window results.

In order to measure latency we control for input throughput, which we fix it at 1 million events per second.
In the fault tolerance experiments of Section~\ref{sec:latency-ft}, we configure Jet to take a state snapshot every second and replicate the snapshots to another 1 member node of the cluster.

\subsection{Latency Under Normal Operation}
\label{sec:latency-nexmark}

\begin{figure*}[t]
    \centering
    \includegraphics[width=\textwidth]{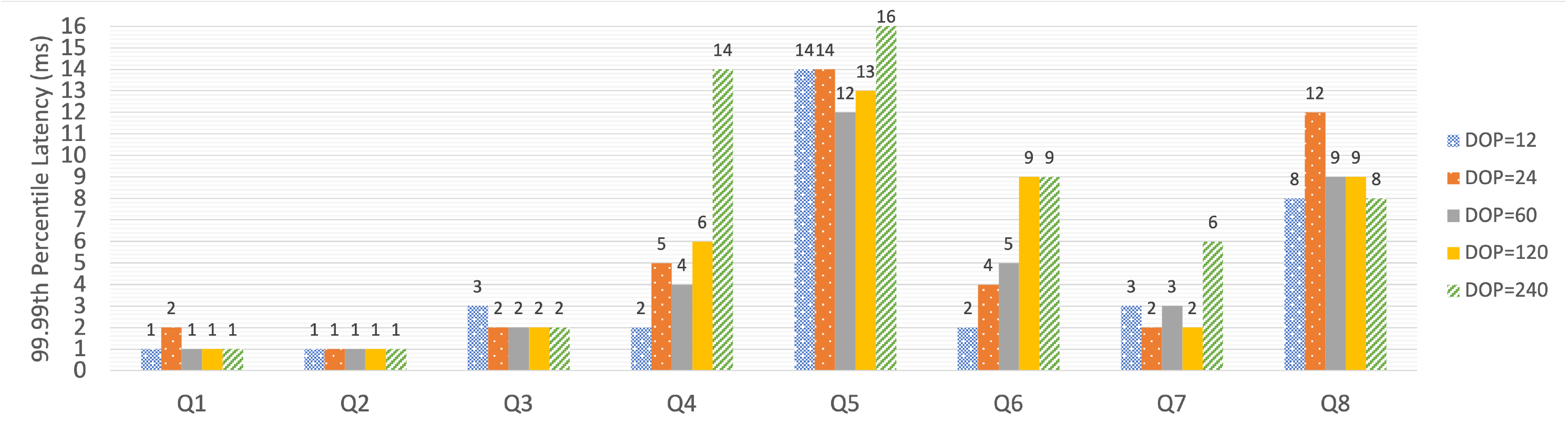}
    \vspace{-4mm}
    \caption{99th percentile latency for all NEXMark queries for fixed input throughput of 1M events/s.}
\label{fig:latency-nexmark}
\end{figure*}

\begin{figure}[t]
\begin{center}
\includegraphics[width=\columnwidth]{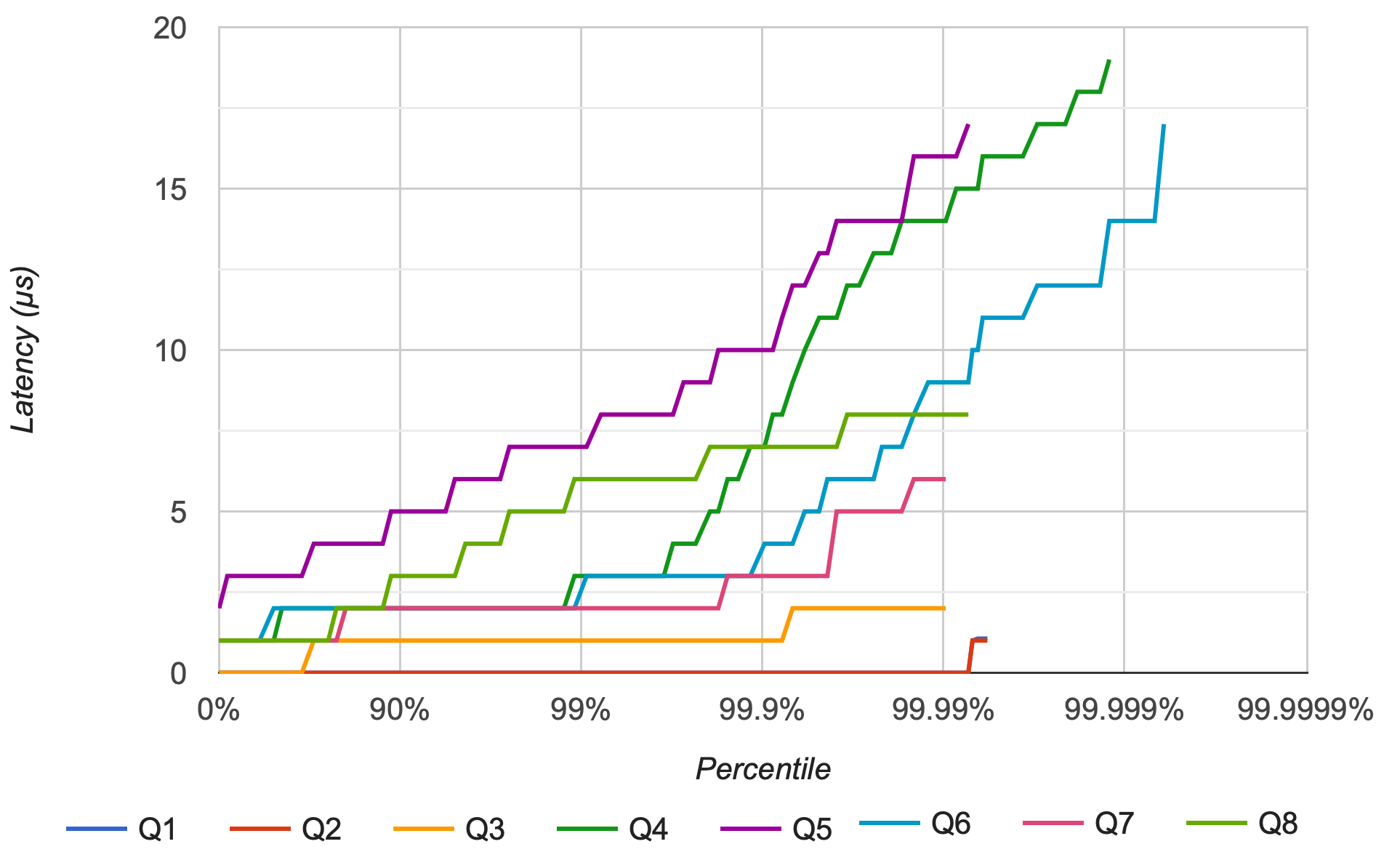}
\caption{Distribution of latencies of all NEXMark queries for 1M events per second and cluster size of DOP=240 (20 nodes).}
\vspace{-2mm}
\label{fig:latencies-detailed}
\vspace{-4mm}
\end{center}
\end{figure}

In this experiment we want to investigate the effect of scaling out to multiple machines (from 1 to 20), while keeping the input rate constant. The results are depicted in \autoref{fig:latency-nexmark}. Overall, we observe that the latency at the 99.99th percentile never exceeds 16ms (in the case of DOP=240 on Q5). Moreover, we observe that Jet adds very little latency to simple queries (e.g., filters, etc.) while the most challenging one seems to be Q5 and Q8. Moreover, in \autoref{fig:latencies-detailed} we depict the distribution of latencies. We observe that 99.9th percentile latency is, in the worst case 10ms. The low latency, though, is to be expected as the input throughput was set well below the limits of what Jet can sustain. We do increase the throughput for Q5 in the next experiment, in order to see the effect to latency.

\subsection{Throughput vs. Latency: 10ms Slide on Q5}
\label{sec:throughput}

In this experiment we want to push Jet to its limit with respect to throughput per CPU-core on a windowed aggregate of a tiny slide of \emph{10ms} on Query 5.
We deployed Jet on a single node (c5.4xlarge, using 12 out of the 16 vCPUs) and used the size of the key set to vary the total input+output throughput. One sliding window results consists of many key-value pairs, so the output throughput scales linearly with the key set size. We increased the total throughput gradually: from less than half a million items per second per CPU core, up to 2 million events per second per core. Our goal is to establish the maximum throughput that the system can sustain and the effect of throughput on the latency of the system. \autoref{fig:throughput-per-core} depicts the results. What we observe is that when throughput is relatively low (around half a million events per second) Jet can sustain the system's latency very well: around 13ms on the 99.99$^{th}$ percentile. When pushed further, a single Jet processor on a single CPU core can sustain up to 2M events per second raising the 99.99$^{th}$ percentile latency up to 98ms. The take away message from this experiment is that for windowed aggregates, more than 1.75M events per second can put a burden on Jet's processors, increasing its latency quite considerably.

\subsection{Throughput: 500ms Slide on Q5}
\label{sec:maxthroughput}

\begin{figure}[t]
\begin{center}
\includegraphics[width=\columnwidth]{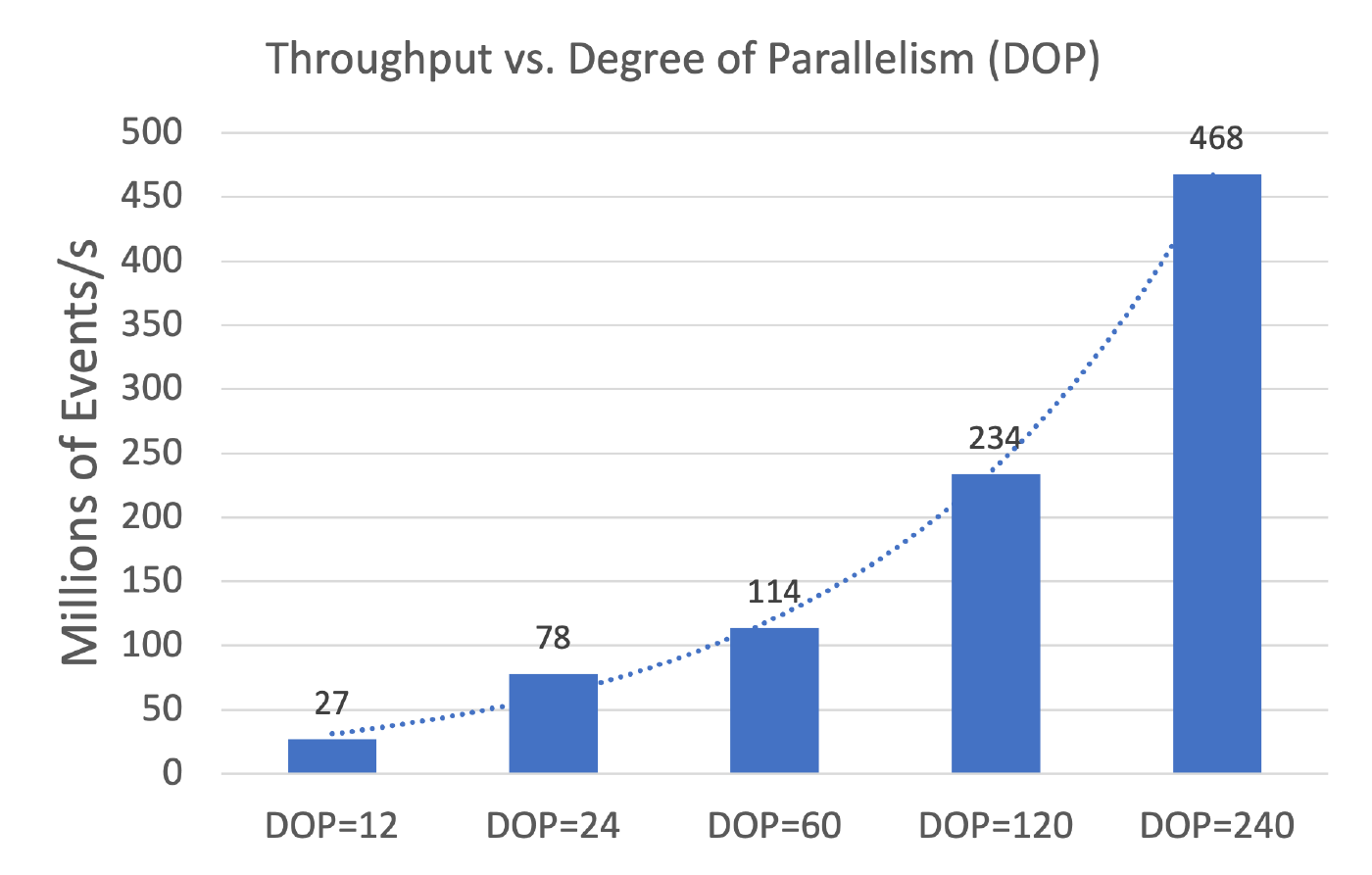}
\caption{Throughput as we increase the cluster size from one VM (12 cores) to 20 VMs (240 cores), for Q5 with a sliding window of 500ms.}
\vspace{-2mm}
\label{fig:throughput-cluster-size}
\vspace{-4mm}
\end{center}
\end{figure}

In this experiment we wanted to evaluate Jet's ability to ingest massive data streams. To this end, we incrased the slide of the windows in Q5 to 500ms, a very reasonable number for this type of queries. As seen in \autoref{fig:throughput-cluster-size} as we increase the cluster size from 12 cores on one VM to 20VMs and a total of 240 cores, Jet manages to ingest up to 468 million events per second. In fact, this was possible due to the use of combiners as after some point the pre-aggregates on the keys (we are usign 10K keys) reach a maximum number and the data exchanged is constant. At the same time, the 99.99th latency never exceeded 17ms. Note that this is possible due to the large window (500ms); in fact, in the previous experiment (\autoref{fig:throughput-per-core}) with the 10ms sliding window, we have seen a very big increase in the latency, as the throughput increased.

\subsection{Latency: Windowed Aggregates \& Joins}
\label{sec:micro}

In this experiment we want to show the capacity of Jet to compute very low-latency aggregates and windowed joins. To this end, we have disabled fault tolerance: as explained in \autoref{sec:ft-tradeoffs} a lot of Jet users prefer to execute jobs in an active-active fashion since Jet allows them to handle very high-throughput streams.

\begin{figure}[t]
\begin{center}
\includegraphics[width=\columnwidth]{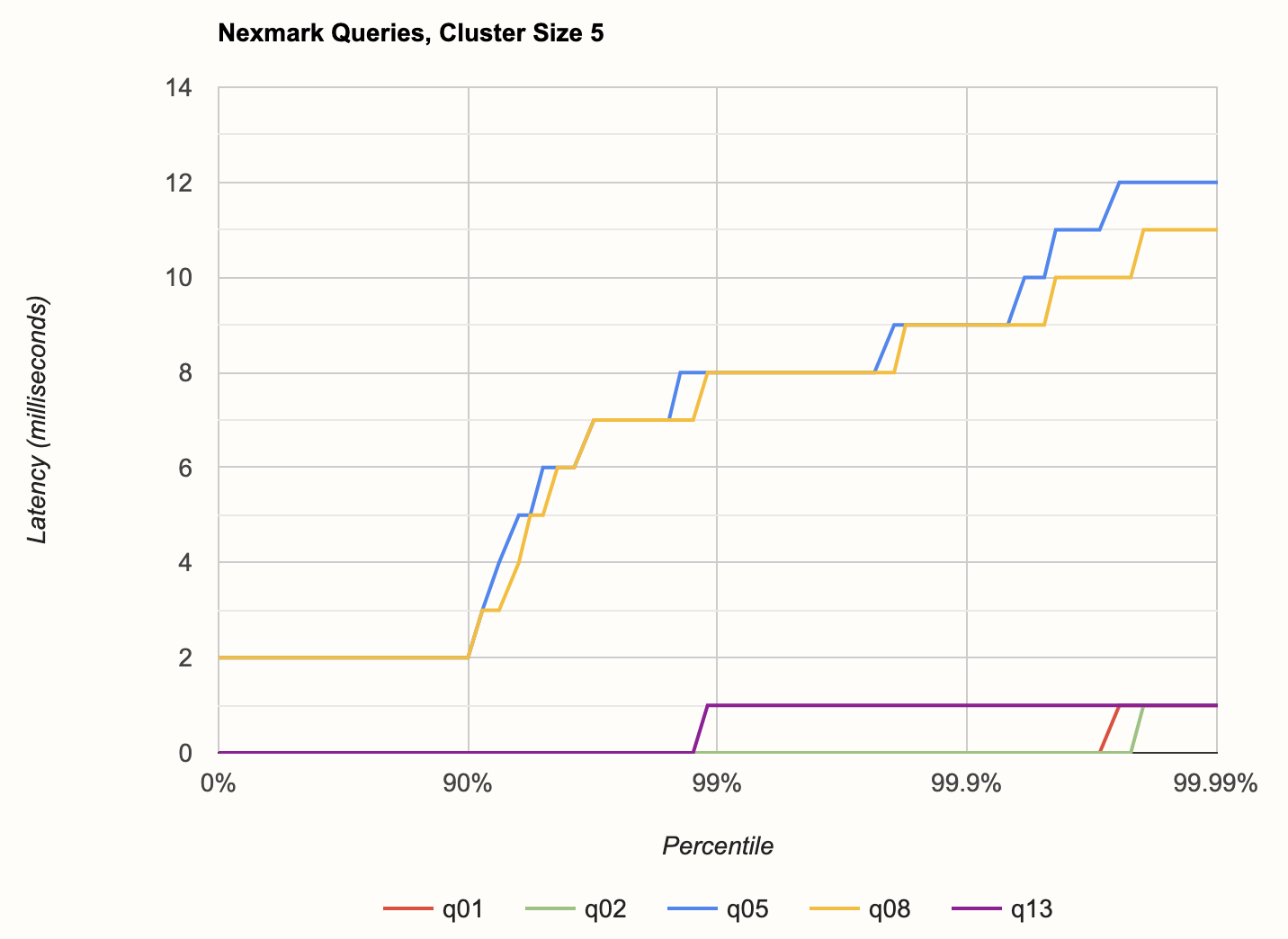}
\caption{Latency for NEXMark queries on a 5-node cluster.}
\label{fig:nexmark-5}
\vspace{-4mm}
\end{center}
\end{figure}

\begin{figure}[t]
\begin{center}
\includegraphics[width=\columnwidth]{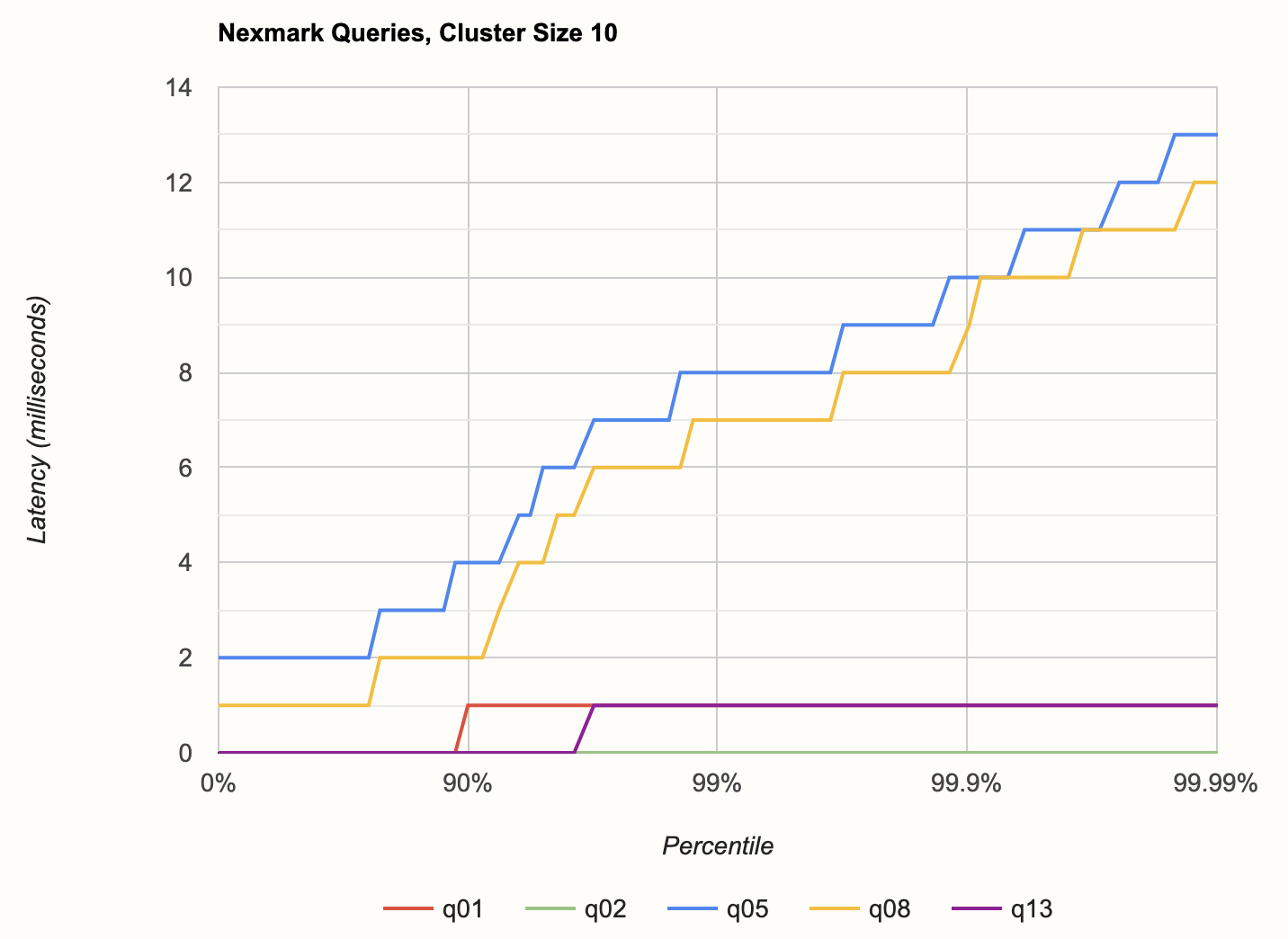}
\caption{Latency for NEXMark queries on a 10-node cluster.}
\label{fig:nexmark-10}
\vspace{-4mm}
\end{center}
\end{figure}

Figures~\ref{fig:nexmark-5} and \ref{fig:nexmark-10} depict the latency for all queries (Queries 1,2,5,8,13) on 5 and 10 nodes respectively. A general observation is that with simple queries that contain only \texttt{map} or \texttt{filter} Jet shows extremely low latency. This is to say that the 99.99$^\text{th}$ percentile stays at or below 1ms. At the same time, queries with joins exhibit 99.99$^\text{th}$ latencies in the order of 11-12ms. Bear in mind that the window triggers every 10ms, meaning that this is an extreme query example that exemplifies the capacity of Jet for low-latency processing. At the same time, more than 90\% of the events exhibit a latency of 2ms or less for joins and equal or less than a millisecond for the windowed aggregates.

\subsection{Latency: Fault Tolerance}
\label{sec:latency-ft}

\begin{figure}[t]
\begin{center}
\includegraphics[width=\columnwidth]{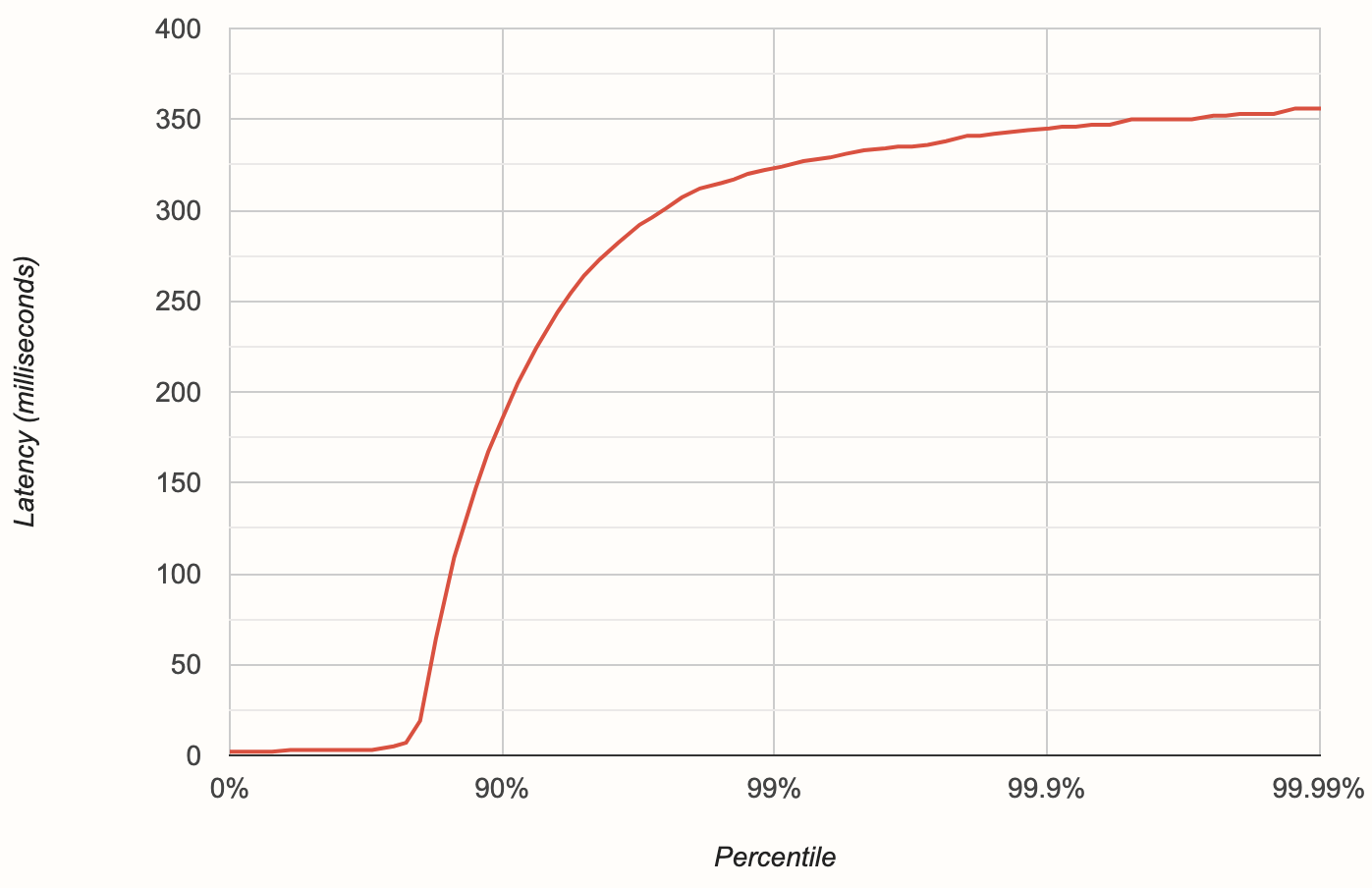}
\caption{Latency in Query 5, with checkpoints enabled.}
\label{fig:ft}
\vspace{-4mm}
\end{center}
\end{figure}

Fault tolerance experiments suggest the overhead of Jet's snapshot mechanism under normal operation. 
As seen in \autoref{fig:ft} Jet's latency at the 99.99$^\text{th}$ percentile when checkpoints are enabled is about 350ms.
Latency remains very low for 70\% of the events approximately, then spikes up to approximately 200ms at the 90\%, and continues to rise sharply up to the 99\%th percentile where it smoothens until it stabilizes at the 99.99th percentile. 

As discussed in \autoref{sec:ft-tradeoffs}, the Jet team has not focused on further optimizing the fault-tolerance mechanism, but rather focused on optimizing the datapath under normal operation. We do have plans on optimizing the datapath with fault-tolerance enabled in the future, especially focusing on at-least once processing guarantees.

\subsection{Latency: Multi-tenancy}
\label{sec:multi-tenancy}

Jet's lightweight execution model allows numerous jobs to execute on the same thread without a considerable performance lag. This is a feature of Jet that comes for free, given its design with tasklets and cooperative threads. In fact, we executed one hundred Query 5 jobs concurrently on a single node, to observe the effect of multi-tenancy to the latency of the system. We observed roughly 200ms 99.99th percentile latency, when running 100 concurrent jobs with an aggregate throughput of one million events per second.

\section{Related Work}
\label{sec:related-work}

A lot of streaming systems have made their debut in the previous years driving important advances.
Some systems, such as Twitter's Storm~\cite{ToshniwalT14} and Heron~\cite{KulkarniBF15}, prioritize performance over consistency, while others like Google's Millwheel~\cite{akidau2013millwheel} and Microsoft's Timestream~\cite{QianHS13} and Streamscope~\cite{LinHZ16} favor results with strict consistency guarantees.
Academia has also developed novel streaming systems such as Apache Spark~\cite{ArmbrustDT18}, Apache Flink~\cite{CarboneKE15}, and Seep~\cite{FernandezMK13}, which have paved new roads in fault tolerance, scalability, and performance.
Notably, open source communities have contributed significantly to the success of such systems.
At the same time, systems like Naiad~\cite{murray2013naiad} and IBM Streams~\cite{SilvaZD16} expand the traditional DAG streaming model of computation to include cycles for instance.
Another class of systems such as Microsoft's Trill~\cite{chandramouli2014trill} inspired a library implementation of a stream processor that can be easily embedded to an application.
Finally, LinkedIn's Samza~\cite{NoghabiPP17} has been built around the concept of a messaging system, i.e. Apache Kafka, which has comfortably found a place in the architecture of modern event-driven applications.
A survey on the evolution of streaming systems with respect to out-of-order data management, state management, fault tolerance and high availability, and elasticity and reconfiguration is provided in reference~\cite{fragkoulis2020survey}.

Hazelcast Jet has certainly been inspired by existing streaming systems. The design of its programming model, out-of-order processing, fault tolerance, and elasticity are notable examples.
On the other hand, because Jet is driven by the need for low-latency performance, it introduces a novel execution model backed by Hazelcast's distributed in-memory grid.
This combination of novel key features sets it apart from existing systems and is validated by the experiments:
Jet provides for reliably high performance that qualifies for strict service level agreements in production use cases.

\section{Event-Driven Applications as Stateful Dataflow Graphs}
\label{sec:future-jet}

Jet aims at becoming an execution platform for event-driven applications~\cite{Katsifodimos2019Operational}, such as microservices~\cite{newman2015Building} and stateful functions~\cite{Sreekanti2020Clouburst, AkhterFK19},
and offer a novel high-level programming model that allows users to convey the business logic of their application by means of standard functions.
In this model, functions can simply call other functions and they can maintain state.
Moreover, functions can be involved in a transaction with other functions while Jet guarantees that changes to the state are applied ensuring atomicity.
Transactions can be marked by programmers with annotations and data flows and dependencies can be automatically derived with static analysis.
Finally, Jet can take care of partitioning and replicating the state while keeping the operational state local to the function it belongs for fast access.

\para{Jet's as a Backend for Functions} The inherent design of Jet offers a number of key features that make it a good candidate platform for running event-driven applications.
First, Jet is particularly good at providing low-latency event processing, and keeps state local to computations.
In addition, the IMDG can be used for fault tolerance, and especially high availability by ensuring that state is partitioned and replicated across the cluster.
Furthermore, as we described in Section~\ref{sec:execution-model}, Jet maintains a pool of threads that corresponds to the available CPUs of a member node.
Since Jet can share a thread among numerous lightweight tasklets, it can overload a thread with tasklets scheduling only those that have pending requests and parking all others until a new request arrives for them.
This design entails that Jet can deal with tens of thousands of tasklets on a single execution thread and take advantage of lightweight statistical multiplexing for e.g., stateful functions \cite{AkhterFK19}.

\para{Open Problems} At the same time, Jet lacks a high-level \textit{application} programming model and lacks support for transactions.
In addition, dynamic reconfiguration, which is required for updating application components like microservices without disrupting the service, is also missing.
Finally, Jet requires a query interface that can perform ad-hoc queries or maintain views over distributed state much like a database system.
This is just a small subset of a long list of open problems that we are currently tackling at Hazelcast.

\section{Conclusions}
\label{sec:conclusions}

In this paper we present Jet, a high performance distributed stream processor built from the grounds up with the goal of minimizing latency at the 99.99th percentile.
Jet has been the favorite choice of customers for a big variety of use cases, such as banking transactions, internet of things, and stateful AI.
We elaborate Jet's execution model and in-memory state backend, the IMDG, which are the key architectural components that make Jet special.
In addition, we describe Jet as a full-fledged streaming system offering fault tolerance with varying consistency guarantees, elasticity and reconfiguration for scale-in scale-out actions, and out-of-order processing.
Experiments with Jet show that it can support service-level agreements on less than 10ms latency at the 99.99th percentile.
We envision the future of Jet as a serverless platform for running event-driven applications, like microservices and stateful functions.


\balance
  
\bibliographystyle{reference-format}
\bibliography{references}
  
  
\end{document}